\documentclass[%
 aip,
 amsmath,amssymb,
 reprint,%
]{revtex4-1}

\usepackage{graphicx}% Include figure files
\usepackage{dcolumn}% Align table columns on decimal point
\usepackage{bm}% bold math
\usepackage{xcolor}
\usepackage[utf8]{inputenc}
\usepackage[T1]{fontenc}
\usepackage{mathptmx}
\usepackage{etoolbox}
\usepackage{siunitx}
\graphicspath{ {./figures/} }
\usepackage{subcaption}
\usepackage{amsmath}
\usepackage{float}
\usepackage[section]{placeins}

\makeatletter
\def\@email#1#2{%
 \endgroup
 \patchcmd{\titleblock@produce}
  {\frontmatter@RRAPformat}
  {\frontmatter@RRAPformat{\produce@RRAP{*#1\href{mailto:#2}{#2}}}\frontmatter@RRAPformat}
  {}{}
}%
\makeatother
\begin{document}

\preprint{AIP/123-QED}

\title[]{Ultrasensitive electrode-free and co-catalyst-free detection of nanomoles per hour hydrogen evolution for the discovery of new photocatalysts}
% Force line breaks with \\
\author{Huaiyu(Hugo) Wang}

\affiliation{ 
Department of Material Science and Engineering, The Pennsylvania State University, University Park, Pennsylvania 16802, USA.
}%
\author{Rebecca Katz}
\affiliation{ 
Department of Chemistry, The Pennsylvania State University, University Park, Pennsylvania 16802, USA.
}%
\author{Julian Fanghanel}
\affiliation{ 
Department of Material Science and Engineering, The Pennsylvania State University, University Park, Pennsylvania 16802, USA.
}%
\author{Raymond E. Schaak}
\affiliation{ 
Department of Chemistry, The Pennsylvania State University, University Park, Pennsylvania 16802, USA.
}%
\author{Venkatraman Gopalan}
 \homepage{Author to whom correspondence should be addressed. Electronic mail: vxg8@psu.edu}
\affiliation{ 
Department of Material Science and Engineering, The Pennsylvania State University, University Park, Pennsylvania 16802, USA.
}%

\date{\today}% It is always \today, today,
             %  but any date may be explicitly specified

\begin{abstract}
High throughput theoretical methods are increasingly used to identify promising photocatalytic materials for hydrogen generation from water as a clean source of energy. While most promising water splitting candidates require co-catalyst loading and electrical biasing, computational costs to predict them \textit{apriori} becomes large. It is therefore important to identify bare, bias-free semiconductor photocatalysts with small initial hydrogen production rates, often in the range of tens of nano-mols per hour, as these can become highly efficient with further co-catalyst loading and biasing. Here we report a sensitive hydrogen detection system suitable for screening new photocatalysts. The hydrogen evolution rate of the prototypical rutile TiO\textsubscript{2} loaded with 0.3 \% wt Pt is detected to be 78.0±0.8 \textmu mol/h/0.04g, comparable with the rates reported in literature. In contrast, sensitivity to an ultralow evolution rate of 11.4±0.3 nmol/h/0.04g is demonstrated for bare polycrystalline TiO\textsubscript{2} without electrical bias. Two candidate photocatalysts, ZnFe\textsubscript{2}O\textsubscript{4} (18.1±0.2  nmol/h/0.04g) and Ca\textsubscript{2}PbO\textsubscript{4} (35.6±0.5 nmol/h/0.04g), without electrical bias or co-catalyst loading, are demonstrated to be potentially superior to bare TiO\textsubscript{2}. This work expands the techniques available for sensitive detection of photocatalytic processes towards much faster screening of new candidate photocatalytic materials in their bare state.
\end{abstract}

\maketitle

\section{\label{sec:level1}INTRODUCTION}

Efficient and clean energy production promises to reduce or reverse the detrimental effects of anthropogenic global warming by reducing carbon emission \cite{RN4,RN5,RN6} and meet the increasing energy consumption needs worldwide \cite{RN7}. Replacement of fossil fuels with various sources of renewable energy requires a cost-efficient method of sustainable production without carbon emission \cite{RN9}, a reliable and low-cost method of energy storage and distribution \cite{RN10}, and a large production volume that can satisfy increasing market needs. Hydrogen is a prime candidate as a renewable fuel \cite{RN12,RN13} and is an efficient energy carrier with an energy density that is three times that of gasoline \cite{RN15}. While conventional methods to produce hydrogen mostly involve steam reforming, a process that releases carbon dioxide \cite{RN25}, photocatalysis offers a potential solution for carbon-neutral generation of hydrogen by electrochemically cleaving water \cite{RN26}. Given that the earth receives \num{3e-24} J of solar energy annually \cite{RN209}, solar production of hydrogen can potentially contribute significantly to the realization of a totally renewable energy supply that can largely satisfy the primary energy consumption needs of the world.

Direct photoelectrochemical water-splitting was first shown by Fujishima and Honda in 1972 \cite{RN237}. This work has been followed by decades of intense research into photocatalytic water-splitting systems involving various co-catalysis and sacrificial agents \cite{RN3}. Despite these efforts, the photocatalytic production of hydrogen has been limited by the lack of stable and inexpensive materials with solar-to-hydrogen conversion efficiency exceeding the threshold of 10 \% for practical commercial applications \cite{RN238}. Data driven prediction of new photocatalytic materials provides new opportunities for expediting the discovery and development of efficient photocatalysis \cite{RN249}. Most of the reported photocatalysts require co-catalysts, sensitizers or support materials to reach above 0.1 \textmu mol/h hydrogen production rate \cite{RN3}. This is due to the fact that the photoexcited carriers recombine in the bulk before migrating to the surface reaction sites, and loading different materials can help mitigate this process. However, co-catalyst choices and loading condition combinations can increase exponentially when it comes to materials screening. Thus for efficient screening of candidate materials, it is crucial to be able to test a potential new photocatalytic material as-is, without any co-catalyst loadings or electrical bias.  This motivates our present work in building an ultra-sensitive tool for detecting small rates of hydrogen generation. Using this tool, we study the hydrogen evolution rates of two recently identified photocatalyst materials \cite{RN249}.

Producing large amounts of novel but untested catalyst candidates for rapid screening can be challenging and expensive. It has therefore become important to investigate the photoactivity of these materials requiring small quantities (i.e., mg) of the catalysts, thus motivating the development of a photo-reactor with a small volume that can detect very low quantities of product gas. This in turn requires efficient collection and detection of small amounts of hydrogen over long periods of time with minimal leakage loss. Most of the current photochemical water-splitting instruments or systems involve glass containers with rubber O-ring compression fittings and/or plastic tubing fittings \cite{RN240,RN241,RN26}. These setups work fine with efficient hydrogen evolution processes that have production rates of above 0.1 \textmu mol/h, but are not sensitive enough for reactions with hydrogen production rates of 10 nmol/h, which is an order of magnitude lower. Besides, these systems still have relatively large reactor volumes that require larger amounts of photocatalysts to be tested. Recently, a new design involving gas-phase water-splitting systems with a small volume and a sensitive detection using mass spectrometer (MS) demonstrated sensitive detection of small volumes of hydrogen gas with an error bar of ± 25nmol and a hydrogen evolution rate of 50 nmol/h/0.003g \cite{RN243}. The apparatus design requires precise control of gas transfer and sensitive gas detection, which can be expensive. Furthermore, the detection rate is still not sensitive enough to detect a few nmol/h of hydrogen evolution. 
	
In order to investigate the capability of hydrogen production from bare semiconductor compounds without any co-catalyst loading, we have developed a new experimental apparatus for ultra-sensitive detection of sub-nanomole per hour hydrogen production rates. The reactor requires minimal volume of liquid (5-10 mL) and mg quantities of catalyst suspended in the solution. The setup does not require high-cost components and the detection end is monitored by a gas chromatograph (GC) installed with a thermal conductivity detector (TCD), which costs less than a MS. The results from a pair of standard photocatalysts, namely, 0.3 wt\% Pt loaded TiO\textsubscript{2} rutile phase and bare TiO\textsubscript{2} rutile phase are presented to demonstrate the working principles of the setup. Furthermore, the detection of hydrogen from three recently discovered candidates, Ca\textsubscript{2}PbO\textsubscript{4}, ZnFe\textsubscript{2}O\textsubscript{4}, and MgSb\textsubscript{2}O\textsubscript{6} \cite{RN249}, are presented to showcase the application in fast screening of predicted samples from first-principle calculation without co-catalyst or electrode. While the photocatalytic activity of ZnFe\textsubscript{2}O\textsubscript{4} has recently been investigated and optimized \cite{RN103,RN104,RN105}, our literature search did not reveal previous experimental evidence of the photocatalytic activity of Ca\textsubscript{2}PbO\textsubscript{4}. Of the three candidates, Ca\textsubscript{2}PbO\textsubscript{4} and ZnFe\textsubscript{2}O\textsubscript{4} are confirmed to be active photocatalysts, while MgSb\textsubscript{2}O\textsubscript{6} is shown to be inactive and acts as a control sample for baseline detection sensitivity. To compare the new candidates with standard photocatalysts directly, bare polycrystalline TiO\textsubscript{2} rutile phase is tested and shown to have a smaller hydrogen production rate as compared with bare Ca\textsubscript{2}PbO\textsubscript{4} and ZnFe\textsubscript{2}O\textsubscript{4}. The results motivate further studies of these two photocatalysts, including the development of appropriate co-catalysts.

\section{Materials and Methods}
\subsection{Materials synthesis and preparation}
Rutile TiO\textsubscript{2} powder (Alfa Aesar, 99.9\%) is mixed with 0.3wt\% Pt reduced from 1.0 mM of H\textsubscript{2}PtCl\textsubscript{6} (Sigma-Aldrich) and stirred for 2 h at room temperature. A rotary evaporator with a water bath at 323 K was employed to remove the water. The resulting paste was dried at 383 K for 12 h to yield a yellow powder. This powder was calcined at 773 K in air for 3 h to yield the PtO\textsubscript{x}/TiO\textsubscript{2} catalyst precursor, which was reduced in flowing hydrogen at a rate of 30 mL min-1 at 423 K for 3 h, to yield a dark-gray TiO\textsubscript{2}/Pt photocatalysts. All samples were synthesized by finely grinding and pelletizing a mixture of powders using an agate mortar and pestle in the molar ratios described below. The samples were added to an alumina boat and heated in air in a Lindberg/Blue M tube furnace. The samples were heated at 5 \textdegree C/min and held at 400 \textdegree C and 800 \textdegree C for two hours prior to heating to the final temperature indicated for each sample, unless other parameters are explicitly mentioned. The samples were then cooled to room temperature inside the furnace. 

\textit{Synthesis of Ca\textsubscript{2}PbO\textsubscript{4} powder:} CaCO\textsubscript{3} powder (Alfa Aesar, 99.99\%) and PbO powder (Alfa Aesar, 99.999\%) were combined in a 2:1 molar ratio of CaCO\textsubscript{3}:PbO and heated to 800 \textdegree C for 26 hours in a Lindberg/Blue M tube furnace. Note that the PbO used to produce Ca\textsubscript{2}PbO\textsubscript{4} had an orange color, likely due to Pb\textsubscript{2}O\textsubscript{3} impurities; Pb\textsubscript{2}O\textsubscript{3} was necessary for this phase to form in high yield. However, using X-ray diffraction, the impurity phase is undetected/below the detection limit. \textit{Synthesis of ZnFe\textsubscript{2}O\textsubscript{4} powder:} ZnO powder (Sigma Aldrich, 
$\geq$99.0\%) and Fe\textsubscript{2}O\textsubscript{3} powder (Aldrich, catalyst grade) were combined in a 1:1 molar ratio and heated to 900 ◦C for 72 hours in a Lindberg/Blue M tube furnace. \textit{Synthesis of MgSb\textsubscript{2}O\textsubscript{6} powder:} MgO powder (Alfa Aesar, 99+\%) and Sb\textsubscript{2}O\textsubscript{3} powder (Aldrich, $\geq$99.9\%) were combined in a 1:1 molar ratio, pelletized, and heated at 5°C/min and held at 400°C and 800°C for two hours prior to heating to the final temperature of 1300°C for 48 hours in a Mullite tube furnace.

\subsection{Materials characterization}
\textit{X-ray diffraction:} Powder X-ray diffraction (XRD) was performed
on a Malvern PANalytical Empyrean (3rd gen.) X-ray Diffractometer for 2$\theta$ in the range of 20° to 80°. The pellets of each material were ground to powders prior to analysis. Reference XRD patterns were generated from either the Powder Diffraction File
(PDF) card number: Ca2PbO4: PDF 04-008-2917; ZnFe2O4: PDF 04-002-2708; MgSb2O6: PDF 01-080-4590.

\textit{Field Emission Scanning Electron Microscope:} Field Emission Scanning Electron Microscope (FESEM) was performed on an Apero 2 S SEM using a backscattered electron(BSE) sensitive T1 detector under Opti-Plan mode with a voltage of 7KV and a current of 50 pA. The prepared sample is in powder form on top of a carbon tape prior to analysis.

\subsection{High through-put screening processes}
The instrument is used in a recent high throughput study of new photocatalytic candidates for hydrogen generation \cite{RN237}. Here we propose a general strategy in high throughput candidate screening process, as presented in a decision tree chart in Figure S5, showcasing the unique functionality of the instrument. First, the band gap and band edges of the material candidates is calculated from DFT+U method and compared to water splitting redox potentials; Second, the material candidates predicted to show hydrogen evolution under illumination under sunlight are synthesized in powder forms and phase pure state. The synthesized samples undergo Mott-Schottky measurements and UV-Visible spectroscopy (see reference 13) to experimentally probe band edges and band gaps in order to compare with theoretical calculations. Third, the synthesized candidates without any co-catalyst are tested in the \textit{Continuous-Flow mode}: if there is any hydrogen production detected, then we conclude that the sample is hydrogen production active. If there is no hydrogen, the sample is further tested in the \textit{Gas-Accumulation mode}, since the hydrogen production rate could be below the detection limit of the \textit{Continuous-Flow mode}, which is around 1 \textmu mol/h. Finally, the instrument in the \textit{Gas-Accumulation mode} monitors the generated gas for up to 4 days at 12 hours intervals; often during the process, one may observe the hydrogen production rate changes potentially due to new phases formed under light illumination. As shown in the two examples in Figure 3b, we can identify the time at which the hydrogen production rate is no longer linear, and identify the new phases through XRD characterization of the catalysis compounds or inductively coupled plasma mass spectroscopy (ICP-MS) characterization of the trace elements in the liquid.

\subsection{Gas chromatograph (GC) calibration}
The GC needs to be calibrated to an accurate volume of the hydrogen gas produced for accurate recording of quatitative analysis of gas products. The calibration process includes extracting known amount of 5\% hydrogen balanced with argon samples from a tank using precise microliter gas tight syringes (Hamilton 80956 used in this paper). A plot of the gas volume vs. GC peak area is fitted to a linear or quadratic equation (in the case when the injected gas amount is small, the calibration curve is no longer linear). The GC peak area from the tested new photocatalysts are converted to the gas volume following the calibration curve equation. The Limit of Detection (LOD) for hydrogen for TCD detector is 1 ppm V/V using Ar carrier gas and the smallest amount of hydrogen detected in this study is 25 ppm V/V, which is above the LOD.

\subsection{Data analysis and errorbar calculation}

There are two modes presented in this work, which are described in detail in section 4. For the \textit{Continuous-Flow mode}, the volume of hydrogen is calculated in Equation \ref{eq:1}:
\begin{equation} \label{eq:1}
    V_{H_2}=\frac{V_{H_{2-GC}}\times R_{flow}t}{V_{sampling}}
\end{equation}

where V\textsubscript{H\textsubscript{2}} is the actual hydrogen volume generated over the time interval period (30 minutes in this paper); V\textsubscript{H\textsubscript{2-GC}} is the detected hydrogen volume from the GC sampling tube in \textmu L; V\textsubscript{sampling} is the volume of the sampling tube, which is calculated to be 0.02 inch\textsuperscript{3} (or 0.328 mL); R\textsubscript{flow} is the flow rate reading from the flow meter in mL/min; t is the time interval in minutes. The volume in the sampling tube is calculated from the calibration curve of the gas volume vs. the GC peak area. The main error comes from the flow meter reading, which has an accuracy of 1 mL/min with an error range of ±0.5 mL/min. The error bar is calculated considering the accuracy of the flow rate and the standard error of GC readings of 10 samplings, which is presented in TiO\textsubscript{2}/Pt data in Figure 1b. The error bar of hydrogen evolution rate is calculated from standard error in linear regression fitting.

For \textit{Gas-Accumulation mode}, the volume of hydrogen is calculated in Equation \ref{eq:2}:
\begin{equation} \label{eq:2}
    V_{H_2}=\frac{V_{H_{2-GC}}\times(V_{chamber}+V{circulation})}{V_{sampling}}\times f_{correction}
\end{equation}

where V\textsubscript{H\textsubscript{2}} is the actual hydrogen volume generated up after some time (every 12 hours in this paper); V\textsubscript{H\textsubscript{2-GC}} is the detected hydrogen volume from the GC sampling tube in \textmu L; V\textsubscript{sampling} is the volume of the sampling tube; V\textsubscript{chamber} is the volume of the reaction chamber, which is calculated to be 10 inch\textsuperscript{3} or 164 mL; V\textsubscript{circulation} is the volume of the circulation tubing, which is calculated to be 0.9 inch\textsuperscript{3} or 15 mL; f\textsubscript{correction} is the correction factor accounting for gas sample loss at each detection step due to the purging of the circulation tubing, which is \begin{equation}\frac{(V_{chamber}+V_{circulation})}{V_{chamber}}  =1.09\end{equation}The error of the detection from this mode is mainly from the sampling gas not completely evenly mixed in the Argon gas. The error bar of detected hydrogen volume is determined from the standard error of 10 samplings at each time stamp, which is presented in Figure 3b. The error bar of hydrogen evolution rate is calculated from standard error in linear regression fitting.

\section{Description of the apparatus and reference photocatalyst testing}

\begin{figure*}
\includegraphics[width=1\textwidth]{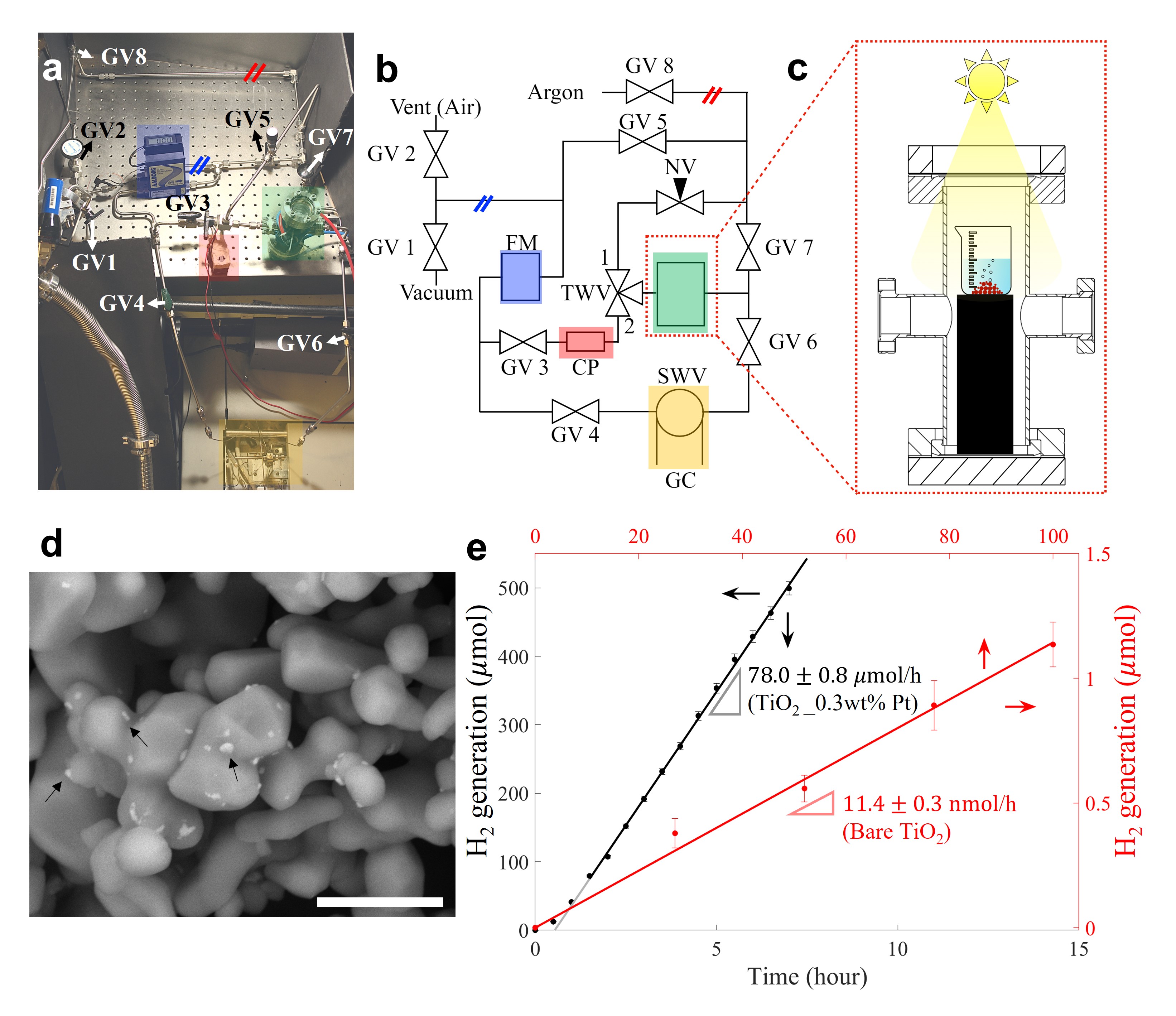}% Here is how to import EPS art
\caption{\label{fig:wide}a) Photograph of the experimental setup. The highlighted region indicates key components and connections in the setup corresponding to the illustration in panel b). b) Illustration of the closed-cycle electrode-free gas chromatography photochemical cell setup. GV: general valve; TWV: three-way valve; FM: flow meter; CP: circulation pump; SWV: six-way valve; GC: gas chromatography; NV: Needle valve. c) Detailed illustration of the reaction chamber. The liquid is mixed with the catalyst in the beaker and is raised by a post. A viewport is installed on the top. d) The FESEM BSE image of TiO\textsubscript{2} loaded with Pt used as a reference photocatalyst. The arrows highlight the Pt islands shown with Z contrast. The scale bar is 1 \textmu m. e) Comparison between rutile TiO\textsubscript{2} loaded with 0.3 wt\% Pt cocatalyst and bare TiO\textsubscript{2}. 40mg of rutile phase TiO\textsubscript{2} loaded with and without 0.3 wt\% Pt were immersed in 10mL of 15 V\% methanol in water, which was used as an electron donor sacrificial agent.}
\end{figure*}

As illustrated in the block diagram shown in Figure 1 a-c, the apparatus is a closed cycle system designed with a leakage rate of \num{3.1e-5} Pa$\times$\text{m\textsuperscript{3}/s} (Figure S1). Stainless steel (SS) and copper gaskets have been used in all constructions in order to allow the system to operate with minimal leakage.  SS tubing is connected by Swagelok compression fittings in most parts except the reaction cell, where a ConFlat (CF) fitting is used, and the six-way valve (SWV) of the gas chromatography (GC), where a Valco compression fitting is used.  The reaction cell is a low-cost (a few hundred dollars), low volume (164 mL), commercially available high vacuum 4-way reducer cross from Kurt J. Lesker (C-0275-133), with fused silica viewport (VPZL-275Q) installed on the top, CF flange (DN35CF-DN40CF) on the bottom and CF to Swagelok adaptor (F0133X4SWG) installed on both sides. The reactor requires only milligrams of catalyst suspended in 5-10 mL of solution. The light source is a 300W Xenon arc lamp (6258) from Newport. The circulation pump is from IWAKI air pumps (APN-30GD2-W), which required sealing with Torr seal epoxy (110516) to reduce the leakage (see Supporting information). The flow meter is from AALBORG (GFM17). All needle valves, three-way valves, and general valves are from Swagelok. Gas Chromatography (GC) analyses were carried out with 5890II instrument (Hewlett Packward) with thermal conductivity detector (TCD). The GC column was a stainless-steel molecular sieve 5A packed column with 80/100 mesh using argon as the carrier gas. The initial oven temperature was 50 \textdegree C, followed by heating to the inlet temperature at 120 \textdegree C, and finally the detector was kept at 150 \textdegree C. The solar simulator has a 300 W Xe lamp and AM 1.5 global filter (81094) from Newport. The output power is measured to be 114 mW/cm\textsuperscript{2}, close to 1 sun intensity for simulated solar conditions. 

To benchmark the performance of our home-built instrument, we chose a popular and standard semiconductor used in photocatalysis, titanium dioxide (TiO\textsubscript{2}) \cite{RN100}. Early work involving TiO\textsubscript{2} photoelectrochemical hydrogen production using a Pt metal electrode as cathode and a TiO\textsubscript{2} irradiated with UV light was reported by Fujishima and Honda \cite{RN237}. In 1977, G. N. Schrauzer and T. D. Guth reported the photocatalytic decomposition of H\textsubscript{2}O on powdered TiO\textsubscript{2} photocatalysts loaded with small amounts of Pt or Rh metal particles \cite{RN244}.  This led to the development of the mechanism that, on such TiO\textsubscript{2}/Pt photocatalysts, the photo-induced electrons move to the Pt metal island on the surface where they induce reduction reactions, while photo-induced holes migrate to TiO\textsubscript{2} surface where they induce oxidation reactions \cite{RN101}. The rate of photocatalysis is determined mainly by the competitive rates of recombination of electron-hole pairs and the charge separation and harvest of electron and holes. The band-structure alignment of semiconductors with water decomposition into hydrogen and oxygen reaction potential is a prerequisite for watersplitting reactions, but the kinetics often limits the rate of photocatalysis, which means that the most promising water splitting catalysts require co-catalysts for efficient hydrogen production \cite{RN3}. 
Here we chose the rutile phase of TiO\textsubscript{2} to be our reference photocatalyst, since the rutile phase was reported to split water into hydrogen and oxygen while the anatase phase was reported to only evolve hydrogen \cite{RN1}, which leads to surface deactivation overtime in anatase TiO\textsubscript{2} phase. For efficient hydrogen production, \textit{Continuous-Flow mode} (to be described in greater detail later) is used where argon gas continuously flows through the chamber and is measured by a flow meter. An example experiment shown with rutile phase TiO\textsubscript{2} loaded with 0.3 wt.\% Pt and a sacrificial agent of 15 vol \% of methanol in water is used to improve the hydrogen production rate. The choice of Pt loading amount is based on a previous optimization study\cite{RN1103} of Pt loading and the optimal loading range of Pt on TiO\textsubscript{2} P25 is reported to be around 0.25 wt.\%. The loading process described in the Materials and Methods section is followed. The FESEM image of rutile TiO\textsubscript{2} after loading with Pt, as shown in Figure 1d, indicates uniform distribution of Pt islands on TiO\textsubscript{2} grains. The powder solution mix is exposed to illumination from a mercury arc lamp, providing an effective light beam diameter of 25mm with an averaged intensity of 114 mW/cm\textsuperscript{2}, which is close to 1 sun intensity (100 mW/cm\textsuperscript{2}). As shown in Figure 1e, the hydrogen production rate in the steady state region is measured to be 78.0±0.8 \textmu mol/h/0.04g of catalyst. This number is comparable to the surface area optimized TiO\textsubscript{2} based photocatalyst experiments (136.2 \textmu mol/h/0.1g) and is close to the performance of pristine TiO\textsubscript{2} (76.6 \textmu mol/h/0.1g) reported under similar sacrificial agent conditions \cite{RN2}. After testing platinized TiO\textsubscript{2}, we focus on detecting hydrogen produced from a photocatalyst with very slow kinetics, namely, bare rutile TiO\textsubscript{2}. Here we showcase the second mode of this instrument, the \textit{Gas-Accumulation mode} (to be described in greater detail later), for ultra-sensitive hydrogen detection rates as low as several nmol/h. The \textit{Gas-Accumulation mode} operates with the evolved hydrogen gas closed cycled within the reaction chamber, so that it builds up for up to several days before being detected by the GC. As shown in Figure 1e, the bare rutile phase TiO\textsubscript{2} shows a hydrogen production rate of 11.4±0.3 nmol/h/0.04g with no change of hydrogen production rate after 100 hours. Only a couple of studies \cite{RN247,RN102} have reported a hydrogen evolution rate for polycrystalline TiO\textsubscript{2} (not P25 TiO\textsubscript{2} nanoparticles), which is often below the standard instrument detection limit. For example, one study reported the rate of hydrogen production on bare TiO\textsubscript{2} to be zero within an experimental error bar of ±20 \textmu mol/h/g \cite{RN247}. In contrast, the instrument in Figure 1a has an estimated error bar of ±7.5 nmol/h/g, which is three orders of magnitude more sensitive.

\section{\textit{Continuous-Flow} and \textit{Gas-Accumulation} modes}

Before discussing the testing of new photocatalysts, we first discuss the approach that is taken when measuring an unknown candidate material. There are two modes of operation: \textit{Continuous-Flow mode} and \textit{Gas-Accumulation mode}. A flowchart of the procedures for operating the experimental apparatus is shown in Figure 2a. After cleaning the chamber with methanol and then loading the sample in the reaction chamber, the top view portal is attached and the standard procedure used to tighten all ConFlat flanges is followed, including wearing nitrile gloves and appropriate cleaning. Next, all the air is purged from the chamber and the detection lines so that the sample is under pure argon.  For this, one must start pumping the reaction chamber to create vacuum for a short period of time until one notices the liquid vapor pressure in the sample beaker exceeds the vacuum level and the liquid starts to boil.  In order to expel the air completely from the chamber, we use the purging method shown in Figure 2b where the Argon gas flows through the chamber continuously for 12 hours. Following this purging, we purge the GC SWV for 10 minutes using a valve configuration shown in Figure 2c.  Using this configuration to purge GC lines, one can detect through GC if there is any trace amount of nitrogen and oxygen from air left in the chamber. To exclude the possibility that hydrogen is produced from the corrosion of the sample materials without light, one should check the gas inside the chamber after the sample is kept in the dark overnight. If there is no corrosion in the dark, the setup is ready for the water-splitting measurement of the sample of interest. 

\begin{figure}
\includegraphics[width=0.51\textwidth]{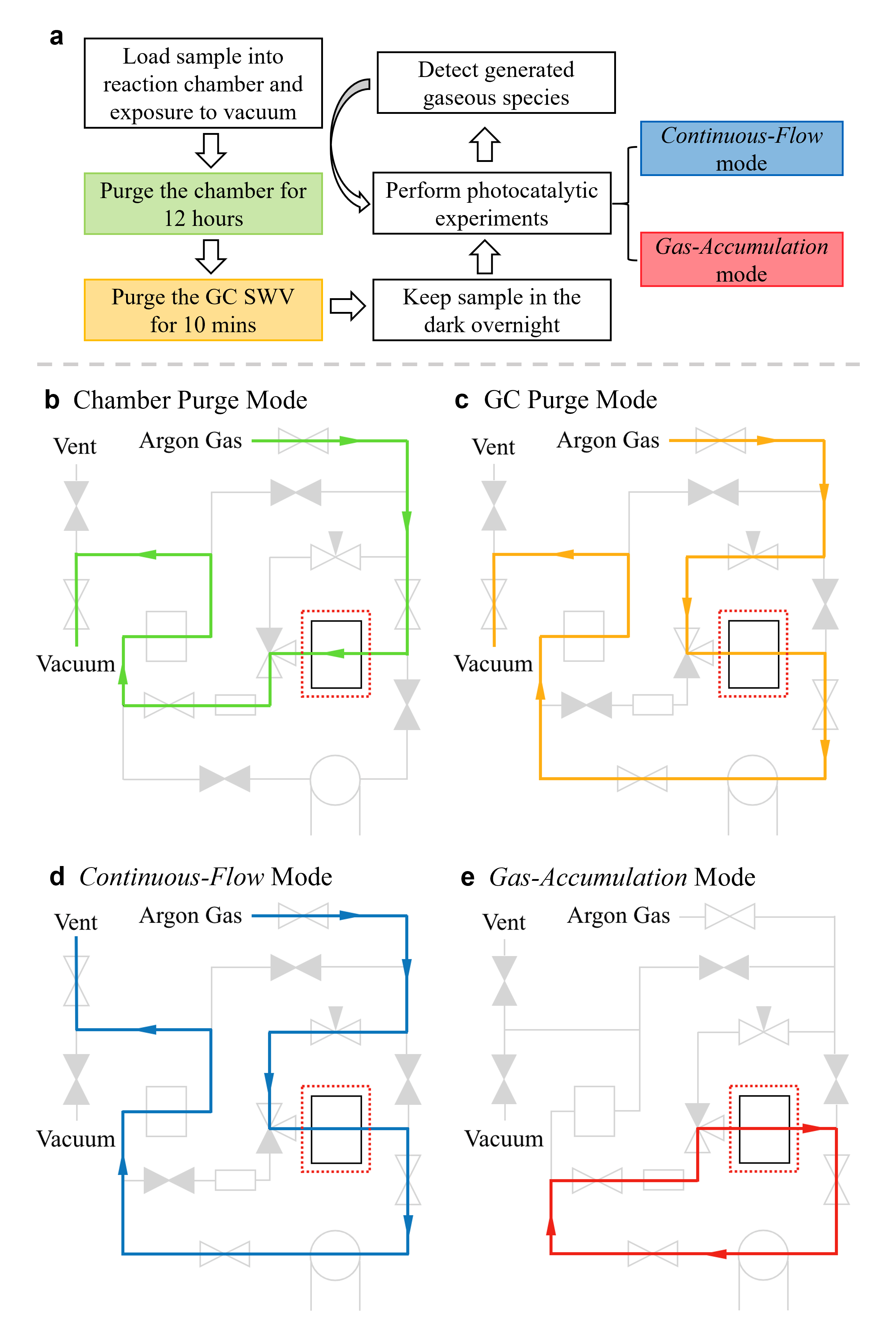}% Here is how to import EPS art
\caption{\label{fig}a) Flowchart of operational procedures. The color codes correspond to setup illustrations in b) to e). Illustration of the operation modes during b) purging the chamber, c) purging the GC SWV, d) \textit{Continuous-Flow mode} and e) \textit{Gas-Accumulation mode}. The colored lines and arrows indicate the flow of argon gas during the operation modes. The grey-filled valve symbols indicates the valve is closed while the unfilled valve symbol indicates the valve is open.}
\end{figure}

One can choose between \textit{Continuous-Flow mode} and \textit{Gas-Accumulation mode} depending on the efficiency of the water splitting process. For \textit{Continuous-Flow mode}, the sample is under a constant flow of Argon gas. The configuration of valves for the \textit{Continuous-Flow Mode} is shown in Figure 2d. During measurement, we found that the vent should be at ambient pressure instead of vacuum, in order to keep a constant flow rate for oxygen and hydrogen. As shown in Figure S2, the correct molar ratio between oxygen and hydrogen from water electrolysis is achieved when the sink is at ambient pressure and a specialized tubing is installed to purge gas right above the liquid level. Once there is a stable flow of argon, gas detection is achieved by switching the GC SWV and collecting the GC spectrum. The speed of the flow is detected using a flow meter and can be controlled either by changing the gas pressure using the gas cylinder valve or adjusting the needle valve. The flow rate should be moderate (approximately above 5mL/min) in order to ensure a steady flow of purge gas. The error bar of the measurement, as shown in TiO\textsubscript{2}/Pt data Figure 1e, is mainly from the flow meter, since its accuracy is 1 mL/min. For \textit{Gas-Accumulation mode}, the photocatalytic water splitting takes place when the chamber is sealed with all valves closed (GV 6,7 and TWV closed) to isolate the circulation pump from the chamber(see discussion in S3). This process is needed to ensure minimum leakage of the gas product during the accumulation time. The configuration of the valves for detection in the \textit{Gas Accumulation mode} is shown in Figure 2e. During detection, the circulation pump is turned on to mix the product gas with the argon gas inside the six-valve sampling tubes. After 2-3 minutes, the circulation pump is turned off. The circulation flow of gas can be monitored by a flow meter to ensure that the gas stops flowing and the pressures in the sampling tube and the chamber are the same before the detection. GC SWV is used to switch the collected gas sample from the sampling tube to flow towards the TCD. This detection process is repeated 10 times to obtain error bar ranges as shown in the pure TiO\textsubscript{2} data in Figure 1e. After each detection step, one must close GV 6 and 7 and TWV (to make sure that the reaction chamber is sealed) and turn on GV1 to keep the circulation tubing and the SWV under vacuum. This step is to ensure minimum leakage during the photocatalytic process. Right before the next detection step, the circulation tubing and the SWV needs to be purged by first pulling vacuum followed by injecting Argon gas by turning GV5,8 on and GV1 off. The purging process is repeated 3 times and no air is detected from GC spectra.

\section{Promising new photocatalysts}
The two operational modes available in this instrument have been demonstrated with bare TiO\textsubscript{2} and TiO\textsubscript{2}/Pt (Figure 1e) and showcase the setup sensitive detection of hydrogen generation rate down to several nmol/h and at the same time capable of characterize efficient hydrogen generation process. In this section, we will demonstrate how the setup is capable of screening new materials from theoretical prediction without co-catalyst or electrical biasing following a procedure described in Figure S5. Three new photocatalysts candidates are predicted through high throughput density functional theory (DFT) calculations \cite{RN249} and tested with this instrument. The computational screening process is described in detail elsewhere \cite{RN249}. Powder samples of the three compounds were synthesized \textit{via} solid-state reactions involving the mixing of precursors and calcinating theses at high temperatures for a given amount of time, as described in the Materials and Methods section. Figure 3a shows the normalized experimental X-ray Diffraction (XRD) pattern, which closely match the reference patterns in all cases, confirming the synthesis of the expected single phase. We then proceed to the photocatalytic activity characterization of these compounds. 

\begin{figure}
\includegraphics[width=0.5\textwidth]{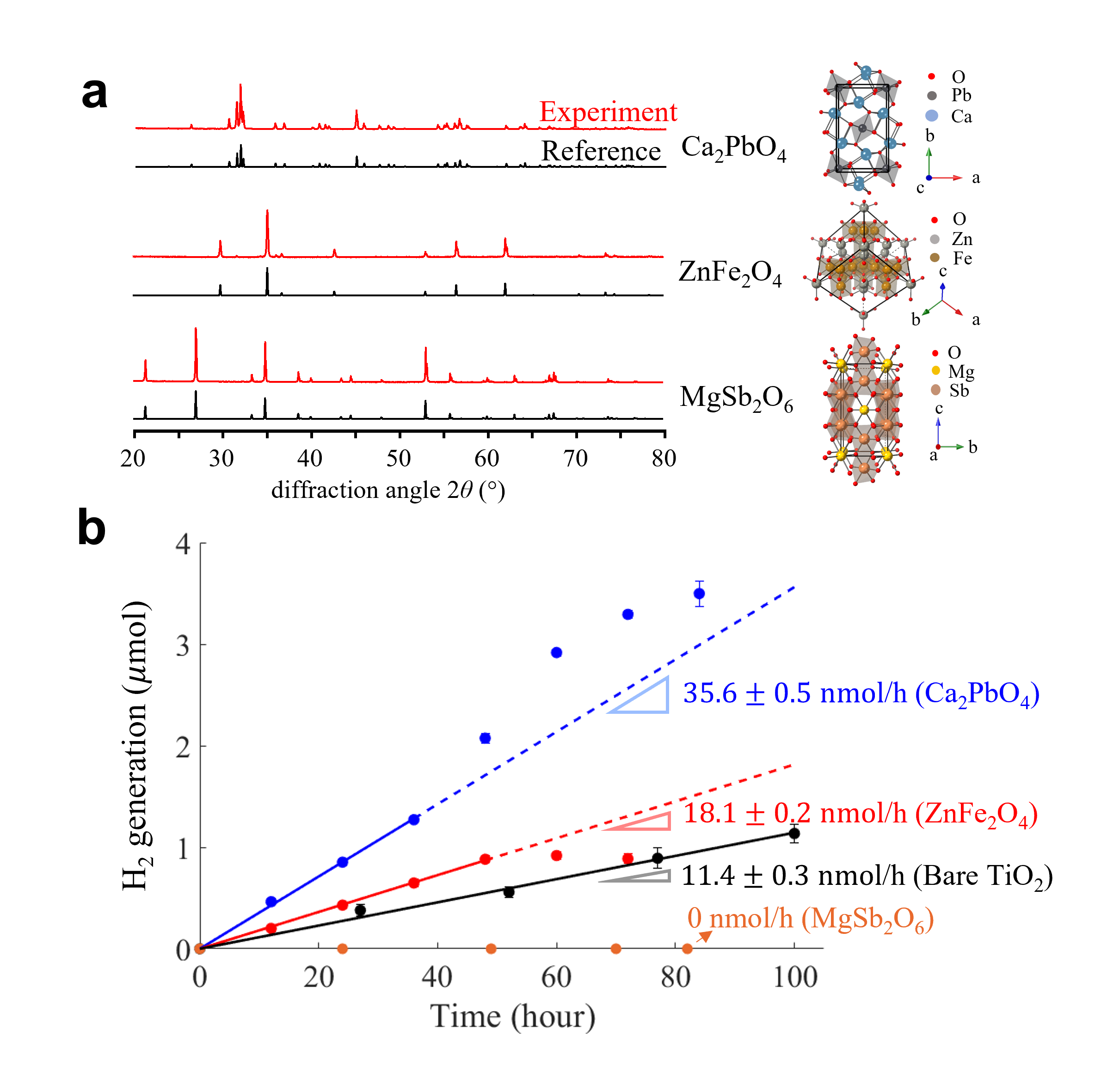}
\caption{\label{fig}a) Comparison of the reference and measured XRD patterns for the 3 compounds that were synthesized and tested. The plane indexing of the peaks is shown in Figure S3. b) Hydrogen evolution test using \textit{Gas-Accumulation mode}. 40mg of Ca\textsubscript{2}PbO\textsubscript{4}, MgSb\textsubscript{2}O\textsubscript{6} and TiO\textsubscript{2} were tested in 10 mL 15 vol\% methanol in water, and 40mg of ZnFe\textsubscript{2}O\textsubscript{4} was tested in 10 mL 0.05 M oxalic acid. Closed dots are experimental data and solid lines are the linear fitting results. The dotted line is an extrapolation of the model that deviates from experimental data indicating corrrosion. The error bar calculation is explained in the Materials and Methods section.}

\end{figure}

To characterize the photocatalytic activity during water-splitting reaction, we implemented the \textit{Gas-Accumulation mode} to measure hydrogen photo-generation. In analysing these measurements, it must be taken into account that the oxygen evolution reaction is much more sluggish than the hydrogen reduction reaction, and often requires loading a cocatalylst to proceed \cite{RN1102}. Although understanding the influence of cocatalysts on the photoactivity is of practical interest for optimizing solar-to-hydrogen conversion, this objective is beyond the scope of the present assessment whose goal is to screen the computationally predicted candidates for solar production of hydrogen. We thus restricted this analysis to the hydrogen reduction half-reaction by introducing sacrificial redox couples to circumvent the slow kinetics of oxygen evolution. The main results expected from the photoactivity tests are two-fold: 1) confirm the location of conduction bands aligned with hydrogen reduction half-reaction and 2) study the corrosion mechanism and how photoactivity is affected by the corrosion process.

Each compound is tested with electron donor agents: acidic pH, with the addition of 0.1 M of oxalic acid, which tends to favor the generation of H\textsubscript{2} (by increasing the activity of the protons) but may also cause the premature dissolution of the sample; (ii) neutral pH, corresponding to volume fractions of 15\% of methanol and 85\% of water. The powder solution mix is exposed to illumination from a mercury arc lamp, providing an effective light beam diameter of 25mm with an averaged intensity of 114 mW/cm\textsuperscript{2}, which is close to 1 sun intensity (100 mW/cm\textsuperscript{2}). From the hydrogen evolution plot shown in Figure 3b, we can conclude that Ca\textsubscript{2}PbO\textsubscript{4} shows an initial hydrogen production rate of 35.6±0.5 nmol/h with a higher hydrogen production rate after 36 hours. Similarly, ZnFe\textsubscript{2}O\textsubscript{4} shows a hydrogen production rate of 18.1±0.2 nmol/h, but becomes inactive after 48 hours. MgSb\textsubscript{2}O\textsubscript{6} shows no hydrogen production and serves as a control test for null detection. In comparison, the bare rutile phase TiO\textsubscript{2} shows a hydrogen production rate of 11.4±0.3 nmol/h with no change in hydrogen production rate after 100 hours. It is noted that although TiO\textsubscript{2}/Pt is a successful and well-studied photocatalyst system \cite{RN2}, the bare TiO\textsubscript{2} sample with sluggish oxidation pathway performs worse than the two new candidates shown here in terms of hydrogen production. On the other hand, TiO\textsubscript{2} evolves hydrogen steadily over 100 hours, indicating no corrosion of TiO\textsubscript{2} happened in PH=7 condition under illumination for 100 hours. 

To further explore the corrosion process during photocatalytic reaction, samples during different hydrogen generation rate period are characterized with XRD to probe the phases. The goal here is to correlate chemical composition of the tested compounds to photoactivity tested over time. As shown in XRD phase analysis in Figure S4a, ZnFe\textsubscript{2}O\textsubscript{4} after 24 hours and 72 hours of photoactivity test indicate the compound decomposes to Fe(C\textsubscript{2}O\textsubscript{4})(H\textsubscript{2}O)\textsubscript{2} in 0.1M oxalic acid under illumination overtime. Combining the XRD characterization with GC test, we can conclude that the corroded compound of ZnFe\textsubscript{2}O\textsubscript{4} is no longer capable of generating hydrogen. On the other hand, the XRD structure characterization of the tested Ca\textsubscript{2}PbO\textsubscript{4} catalyst shown in Figure S4b indicates that  Pb\textsubscript{3}O\textsubscript{4} and CaPbO\textsubscript{3} form after 85 hours of photocatalytic reaction, and they are not present after 24 hours. This suggests that the formation of Pb\textsubscript{3}O\textsubscript{4} and CaPbO\textsubscript{3} are related to the higher hydrogen production measured from Ca\textsubscript{2}PbO\textsubscript{4} after 36 hours. 

\begin{figure}
\includegraphics[width=0.5\textwidth]{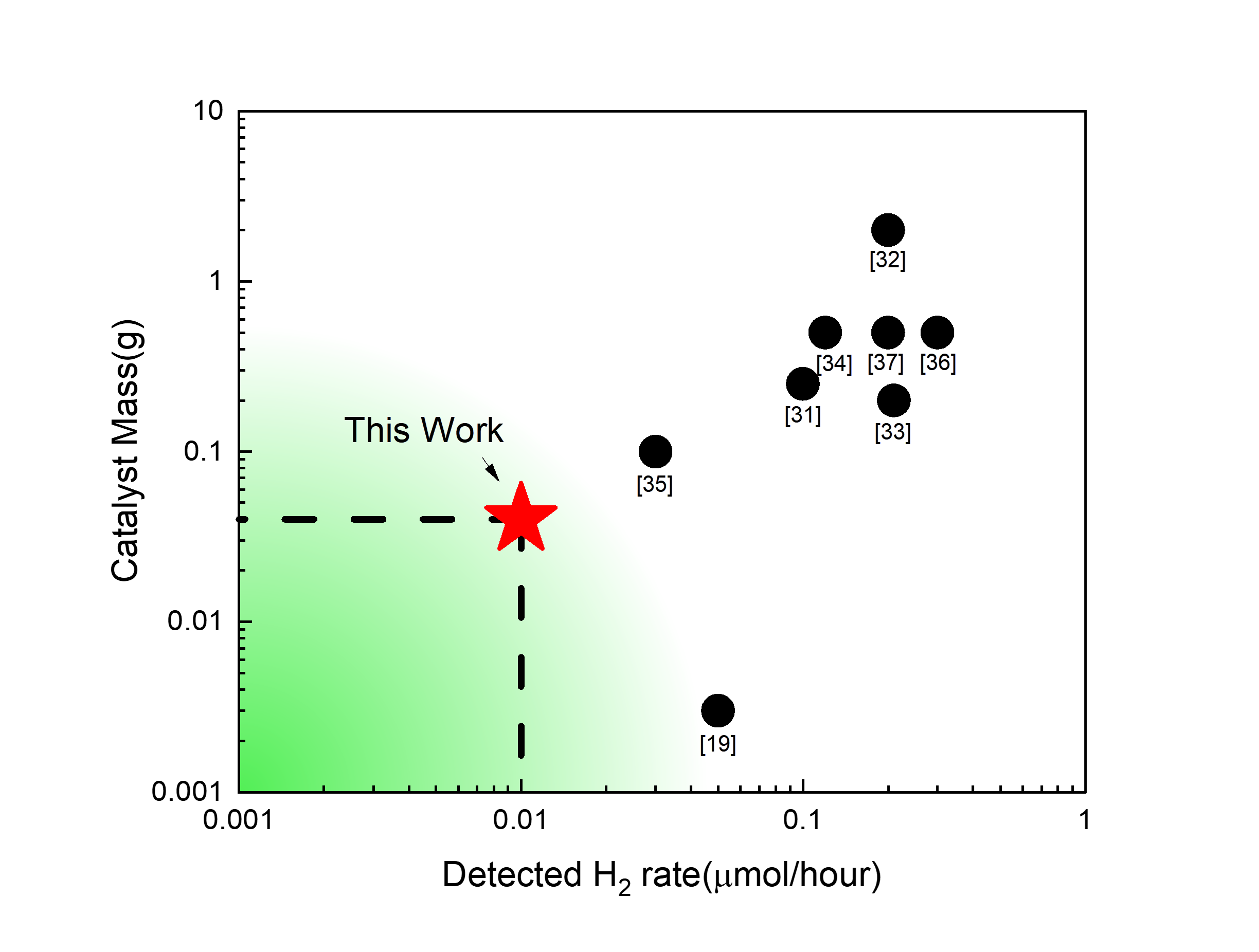}% Here is how to import EPS art
    \caption{\label{fig}Comparison of reported lowest hydrogen detection rate and catalyst mass used from photo activities measurement setups in literature to this work  \cite{RN243,RN1001,RN1002,RN1003,RN1004,RN1005,RN1006,RN1007}. The green region indicating high sensitivity low mass detection desired for photoactivity screening of co-catalyst-free and electrode-free materials}
\end{figure}

To benchmark the performance of our sensitive hydrogen detection system, we choose two important parameters that are important for high throughput screening of photocatalytic materials: 1) sensitive detection of small hydrogen generation rate, and 2) small catalyst mass required for the test to expand the material candidates to those with small quantity that are hard to scale up in the synthesis process. As shown in Figure 4, after extracting reported literature values on photoactivity test with hydrogen production rate of 1 \textmu mol/hour or lower, it is noted that most of literature values cluster above 0.1 \textmu mol/hour and above 0.1 g. The detection setup designs reported in literature are a combination of closed container equipped with rubber septum\cite{RN1005,RN1007}, closed cycled system with Pyrex glass container linked to GC \cite{RN1003,RN1004,RN1006} or MS \cite{RN1001}, or closed container with water vapor reduction \cite{RN1002}. A closed container with rubber septum combined with syringe sampling can achieve sensitive detection level \cite{RN1005}; however the reported hydrogen evolution curve is nonlinear due to a change of partial pressure in the reaction cell after each sampling. Recently reported new design involving gas-phase water-splitting systems requires a few mg of samples and can achieve a sensitive detection of 50 nmol/h/0.003g using MS \cite{RN243}; however, the apparatus design requires precise control of gas transfer, sensitive gas detection and specialized sample preparation. Overall, our design of closed-cycled online hydrogen generation detection system shows superior performance as compared to the above-mentioned sensitive setups from the literature. 

\section{Conclusion}

We demonstrate a specially designed closed cycle experimental setup for the ultrasensitive detection of gaseous products from photocatalysis covering reaction rate over 4 order of magnitude from hundreds of {\textmu}mol/h down to tens of nmol/h and catalyst mass as small as 0.04 g. We demonstrate its utility by studying two photocatalysts, Ca\textsubscript{2}PbO\textsubscript{4} and ZnFe\textsubscript{2}O\textsubscript{4}, that were recently predicted and experimentally validated, both of which exhibit superior performance to TiO\textsubscript{2} in their bare, co-catalyst free, bias-free state. However, ZnFe\textsubscript{2}O\textsubscript{4} decomposes to Fe(C\textsubscript{2}O\textsubscript{4})(H\textsubscript{2}O)\textsubscript{2} in 0.1M oxalic acid under illumination and terminates the hydrogen generation after 48 hours. Ca\textsubscript{2}PbO\textsubscript{4} decomposes to Pb\textsubscript{3}O\textsubscript{4} and CaPbO\textsubscript{3} and enhanced the hydrogen generation after 36 hours. Future studies on corrosion mechanisms of the two compounds under photo illumination might be able to shed lights on pathways to enhance the stability of two studied compounds. These sensitive experimental measurements are made feasible by the low leakage rate of the instrument combined with a small volume size of the reaction chamber, which allows three orders of magnitude higher detection sensitivity for hydrogen evolution than a conventional gas chromatograph system, while still maintaining high sensitivity to trace products. The small volume of the reactor enables the use of small quantities of a catalytic material, thus avoiding large volumes of expensive catalyst preparation. With further improvement in the vacuum and replacing GC detection with a MS, an order of magnitude improvement in both detected H\textsubscript{2} rate and catalyst mass required. High detection sensitivity can thus fast-track experimental screening of new computationally identified photocatalysts for hydrogen production (see experimental methods for a procedure), as well as a variety of quantitative catalytic studies across a wide range of conditions. 

\section*{supplementary material}

See supplementary material for leakage rate of the apparatus, purging condition test, circulation pump sealing, degradation analysis and a description of the high throughput screening process. 

\section*{Acknowledgments}

This work was supported by the DMREF and INFEWS programs of the National Science Foundation under Grant No. DMREF-1729338. The authors are deeply thankful to I. Dabo for fruitful discussions. 

\section*{Author's Contributions} 
H. Wang and V. Gopalan conceived the idea and designed the experiments. H. Wang wrote the main manuscript text. R. Katz and J. Fanghanel synthesized the compounds used in the experiments. H. Wang and R. Katz conducted characterization and analysis of photochemical stability of the compounds. All the authors reviewed and revised the manuscript. V. Gopalan and R.E. Schaak supervised the work.

\section*{Conflicts of Interest}
The authors declare that there is no conflict of interest regarding the publication of this article.

\section*{Data Availability Statement}

Data available on request from the authors.

\bibliography{aipsamp}% Produces the bibliography via BibTeX.

\end{document}

\title{Ultrasensitive electrode-free and co-catalyst-free detection of nanomoles per hour hydrogen evolution for the discovery of new photocatalysts}

\author{Huaiyu(Hugo) Wang, Rebecca Katz, Julian Fanghanel, Raymond E. Schaak, Venkatraman Gopalan}

\date{}
 
\maketitle
\section*{Supplementary Materials}
\renewcommand{\thefigure}{S\arabic{figure}}
\setcounter{figure}{0}
\subsection*{S1. leakage rate of the apparatus}
The leakage test is performed using the pressure rise method \cite{RN1101} under argon environment. The chamber is initially purged with Argon gas(UHP,99.998\%) and then pumped to lowest pressure. Then the pump is isolated from the setup and the pressure is measured as a function of time as shown in Figure S1.
\begin{figure}[!htb]
    \centering
    \includegraphics[width=0.8\textwidth]{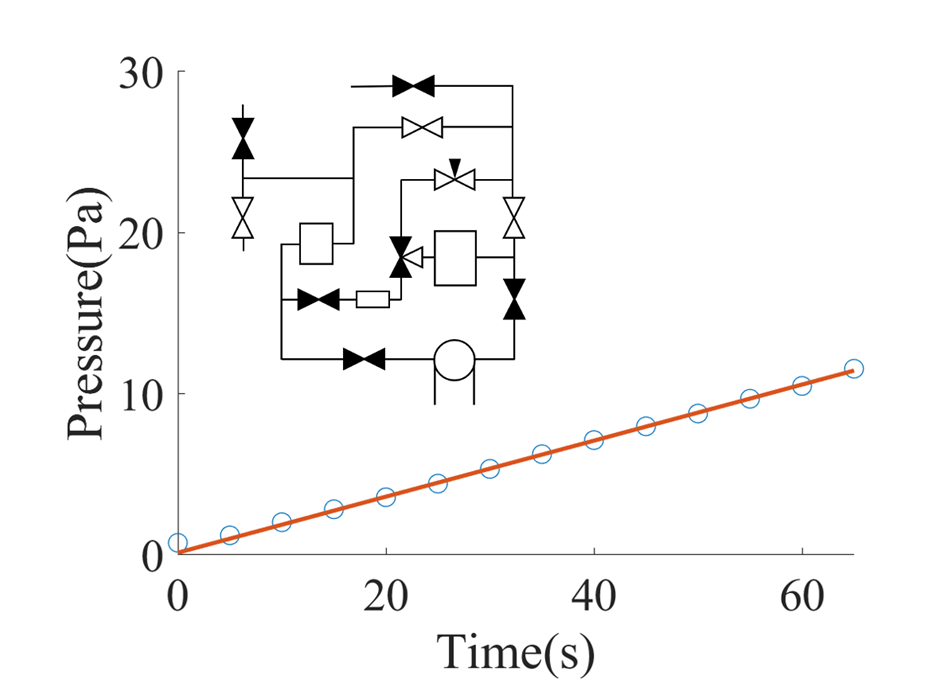}
    \caption{Leakage test result of the photochemical cell setup. Blue dots are recorded pressure data and red line is the linear fitting result. The inset is an illustration of the valve positions during leakage test. The filled valves indicate that they are closed while unfilled valves indicate that the they are open. The leakage rate is calculated to be \num{3.1e-5} Pa$\times$\text{m\textsuperscript{3}/s}.}
    \label{fig:S1}
\end{figure}
\vspace{5mm}
\subsection*{S2. A comparison of full water splitting V\textsubscript{H\textsubscript{2}}/V\textsubscript{O\textsubscript{2}} between Vacuum vent and atmospheric air vent, with and without tubing installation}

\begin{center}
    \includegraphics[width=0.8\textwidth]{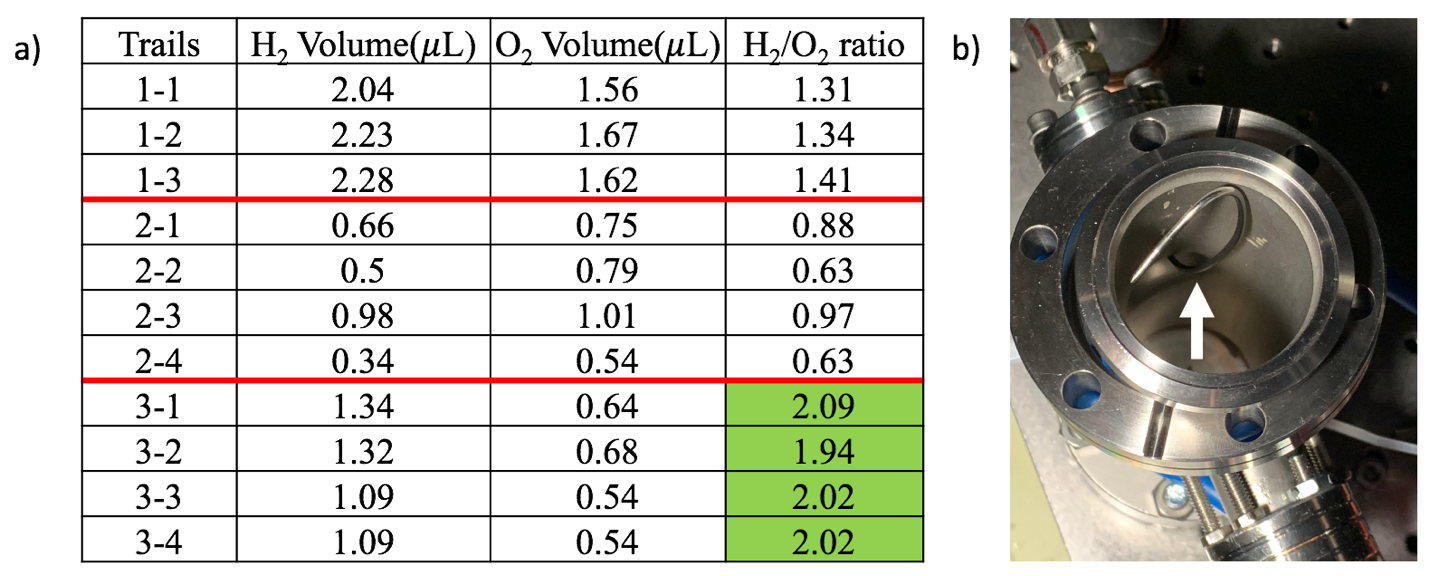}
    \captionof{figure}{A summary of the generated hydrogen and oxygen volumetric ratio of a water electrolysis (full water-splitting process) under different conditions: a) The table summarizes the results of the trail 1): atmospheric air vent without tubing; 2) vacuum vent without tubing; 3) atmospheric air vent with tubing. The highlighted cells in green indicate the correct hydrogen vs. oxygen ratio. b) A photo of the installed tubing, which will be placed right above the liquid surface during the photocatalytic process.} 
    \label{fig:S2}
\end{center}

\subsection*{S3. Sealing of circulation pump for closed cycle circulation }

Due to the design of the commercial circulation pump, air can easily leak into the system via leaks in the pump. To fix these leak spots, Thor seal high vacuum epoxy was applied. The vacuum can reach the highest value (3 mTorr) after the epoxy dries. However, the leakage rate at the circulation pump is still much larger than that at the reaction chamber. As a result, when performing the \textit{Gas-Accumulation mode}, the circulation pump needs to be isolated from the chamber. 
 
 \subsection*{S4. Plane indexing of XRD data}
\begin{center}
    \centering
    \includegraphics[width=1.1\textwidth]{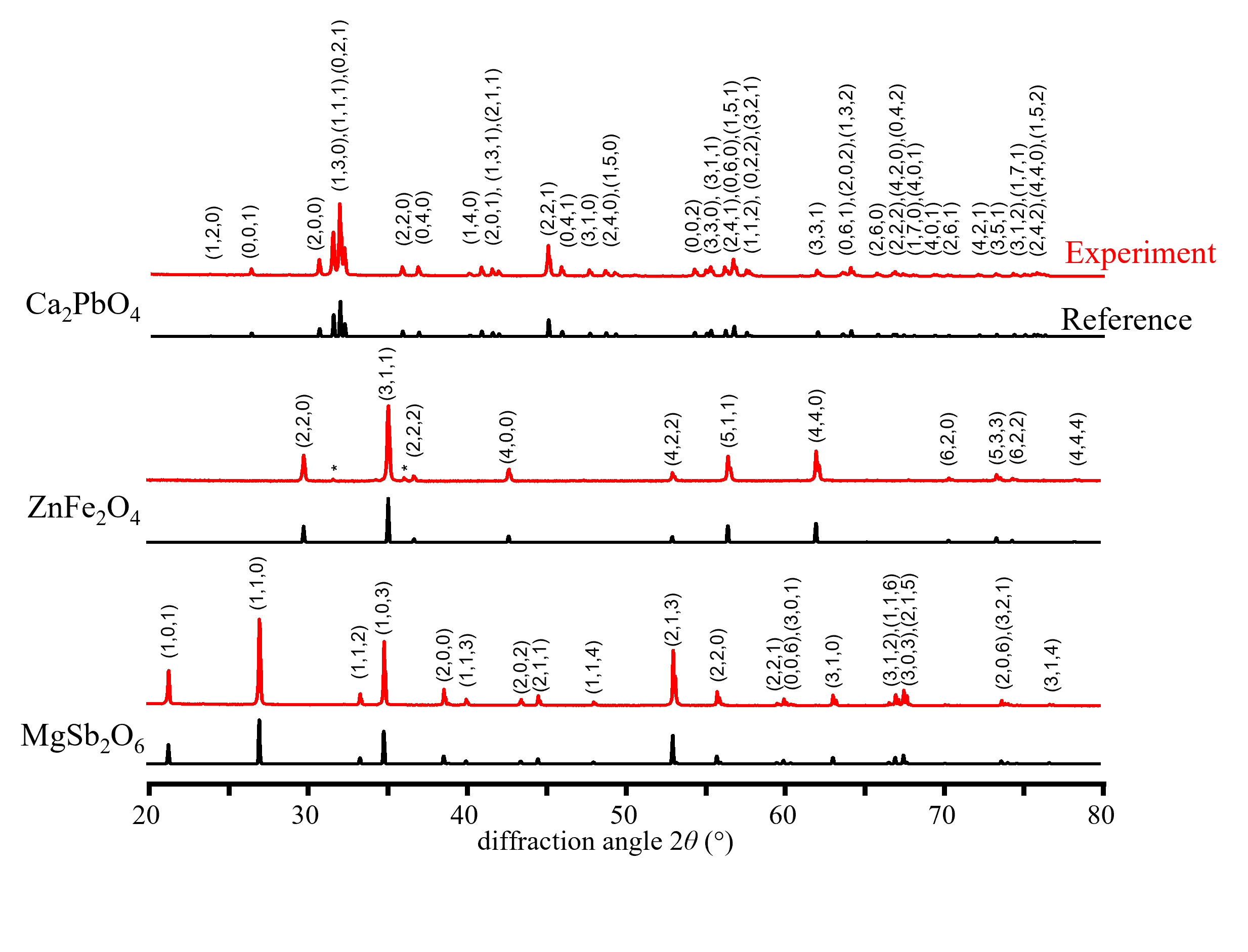}
    \captionof{figure}{Experimental and reference data of XRD results of Ca2PbO4, ZnFe2O4 and MgSb2O6. The * marks the peaks that do not agree with the reference simulation.} 
    \label{fig:S3}
\end{center}
 
\subsection*{S5. Degradation test of the measured samples from XRD}
\begin{center}
    \centering
    \includegraphics[width=1\textwidth]{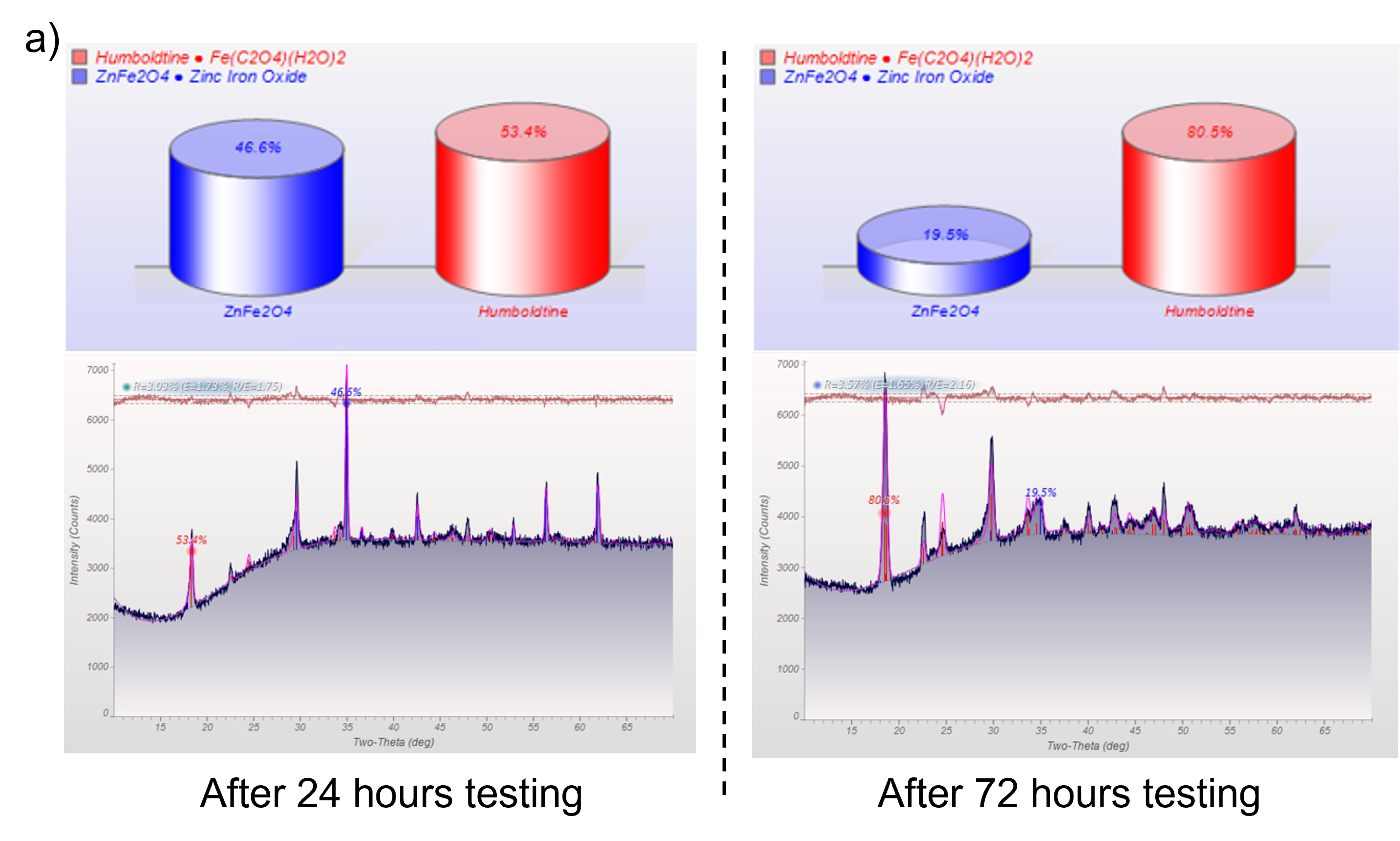}
    \includegraphics[width=1\textwidth]{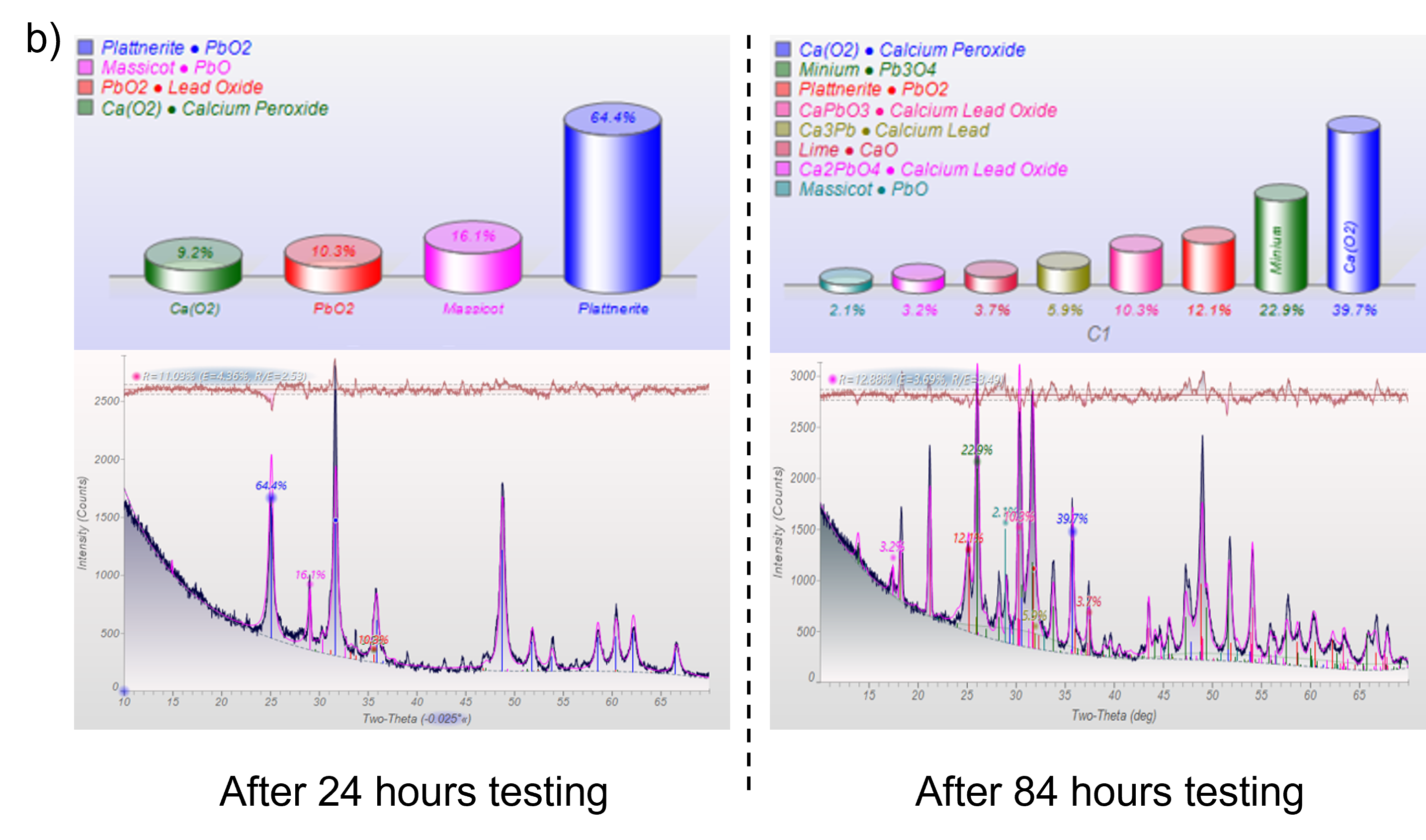}
    \captionof{figure}{XRD characterization and phase analysis results of a) ZnFe2O4 as synthesized, after 24 hours and after 72 hours; b) Ca2PbO4 as synthesized, after 24 hours and after 84 hours.} 
    \label{fig:S4}
\end{center}

\subsection*{S6. High throughput screening processes}
\begin{center}
    \centering
    \includegraphics[width=0.8\textwidth]{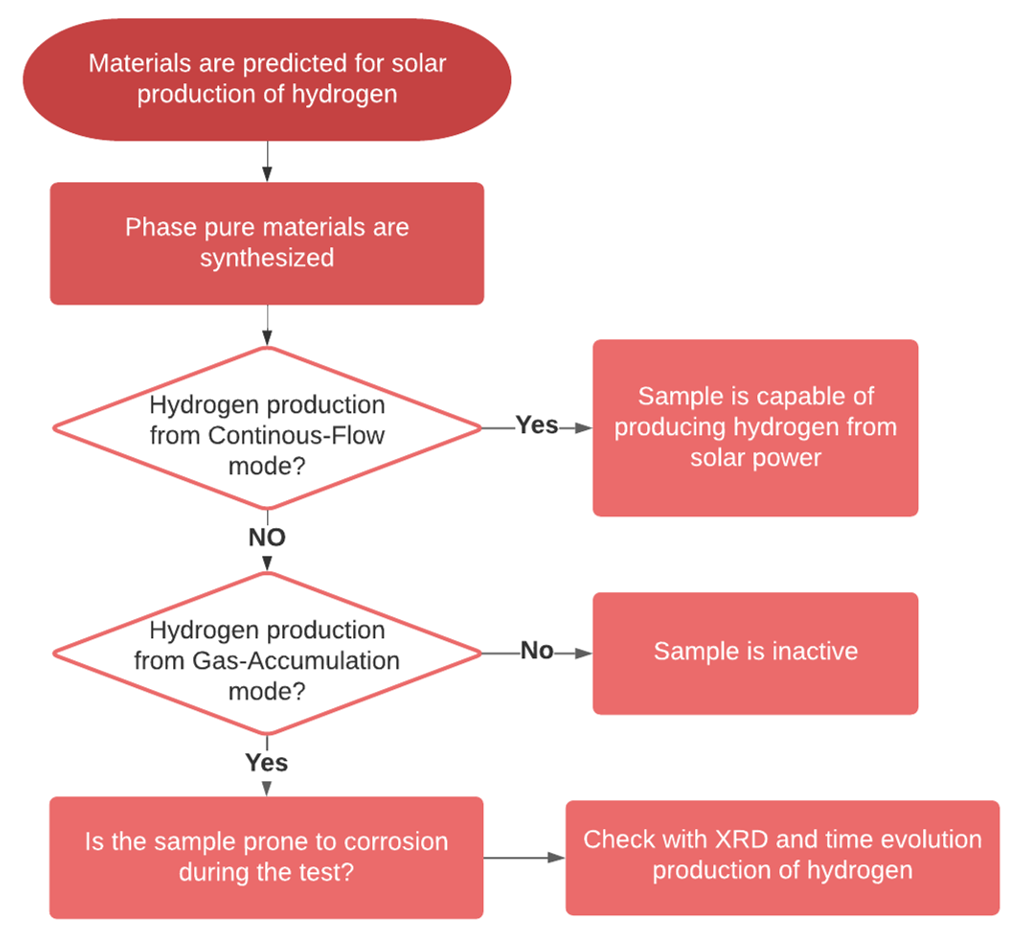}
    \captionof{figure}{Data-driven computational protocols for testing high through-put screening of candidates for solar production of hydrogen.} 
    \label{fig:S5}
\end{center}

\printbibliography